\documentclass[a4paper]{jpconf}
\usepackage{graphicx}
\usepackage{amssymb}
\usepackage{amsbsy}
\usepackage{subfig}
\usepackage{lineno}
\usepackage[utf8]{inputenc}

\begin{document}
\title{Recent results for forward J/$\boldsymbol{\psi}$ production in Pb--Pb Ultra-Peripheral
       Collisions at $\mathbf{ \sqrt{s_{\mathbf  N\mathbf N}} = 5.02 }$ TeV with the ALICE detector}

\author{Simone Ragoni for the ALICE Collaboration}

\address{University of Birmingham, Edgbaston, Birmingham, B15 2TT, United Kingdom}

\ead{simone.ragoni@cern.ch}

\begin{abstract}
The LHC is not only the highest energy collider for protons and heavy ions, but also for photon–photon and photon–hadron ($\gamma$p and $\gamma$Pb) interactions. This is because the protons and ions accelerated in the LHC carry an electromagnetic field, which can be viewed as a source of photons, and such photons can interact with either other photons or with hadrons.
Ultra-Peripheral Collisions (UPC) occur when the incoming ions pass beyond the range of the strong force, with impact parameters larger than the sum of the radii of the incoming projectiles, and are mediated by the exchange of virtual photons between the nuclei. The number of photons scales with the square of the nuclear electric charge, and the photon energies increase rapidly with beam energy. The beam energies at the LHC make the LHC the most energetic photon source ever built. In particular, the photoproduction of heavy vector mesons is favoured because such mesons couple to the photon.
Recent results for forward J/$\psi$ production in Pb--Pb UPCs at $\sqrt{s_{\rm NN}} = 5.02$ TeV with the ALICE detector are presented here.
\end{abstract}

\section{Introduction}
Protons and ions accelerated in the LHC carry an electromagnetic field, hence
they can be viewed as a source of virtual photons. Such photons can interact
with the opposite proton or nucleus, giving rise to ultra-peripheral collisions (UPC)
\cite{doi:10.1146/annurev.nucl.55.090704.151526, 2015IJMPA..3042012C}.

\begin{figure}[ht]
\includegraphics[width=18pc]{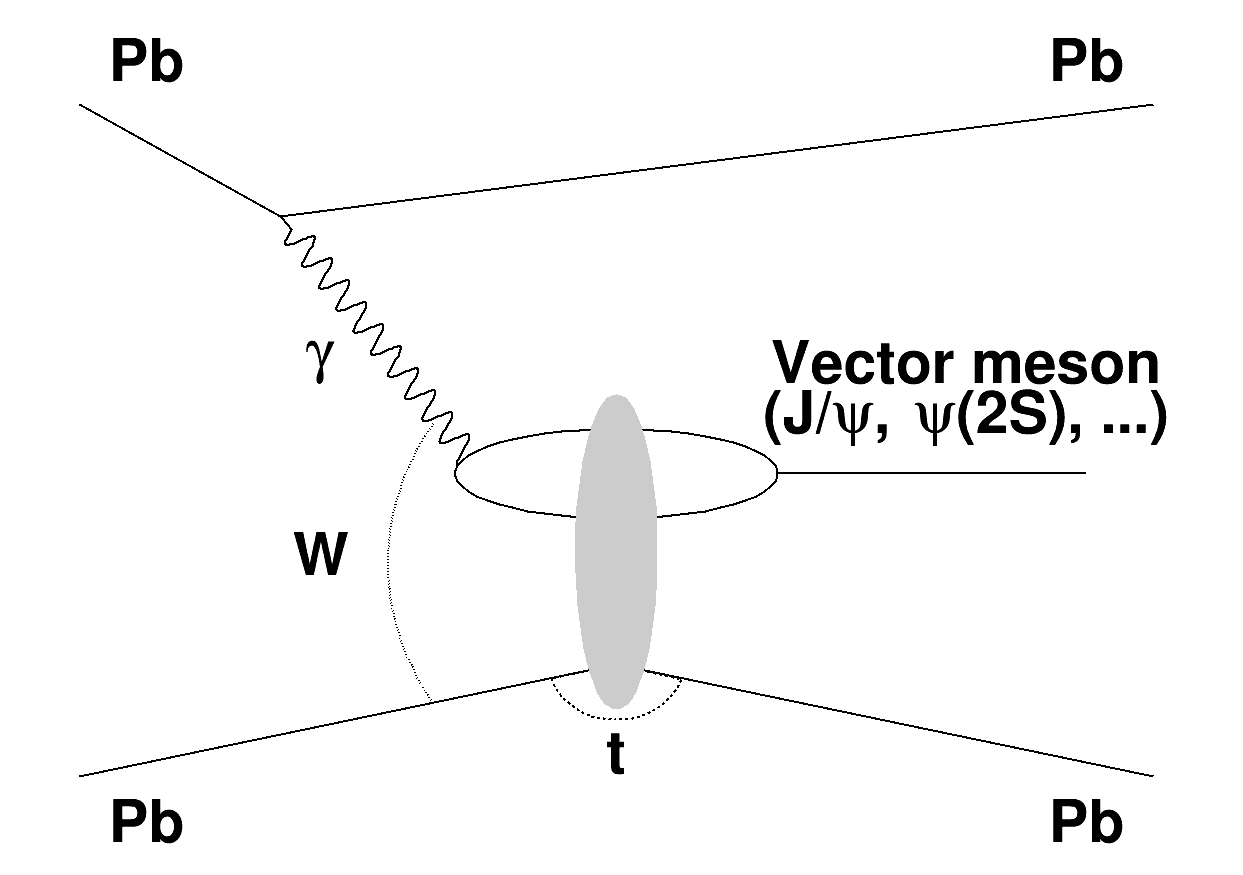}\hspace{2pc}%
\begin{minipage}[b]{14pc}
\caption{\label{VectorMeson} Sketch for vector meson photoproduction in UPC processes.
                             The shaded area represents the interaction of the gluons with the
                             emitted photon, known as a pomeron exchange.}
\end{minipage}
\end{figure}

UPC are characterised by impact parameters $b$ larger than the sum of the two
nuclear radii $R_{\rm A}$ and $R_{\rm B}$, $b > R_{\rm A} + R_{\rm B}$; the photons
involved in the exchange are quasi-real, with a flux proportional to $Z^{2}$, where $Z$
is the charge of the ion or the proton. In the framework of the Generalised
Vector Dominance Model (GVDM) \cite{Sakurai:1960ju, Bauer:1977iq}, the high cross sections involved for this process
can be explained thanks to a possible fluctuation of the $\gamma$ to a $q\bar{q}$
pair. The $\gamma$ has quantum numbers $J^{\rm PC} = 1^{--}$, so vector meson
photoproduction is favoured,  in fact this contribution is focused on
J/$\psi$ photoproduction, and Fig.~\ref{VectorMeson} shows the diagram of such an
event.

These events are sensitive to the gluon distributions in nuclei, as seen in
Fig.~\ref{VectorMeson}. In particular, these processes require a colour singlet
state, hence an exchange of at least two gluons. The low multiplicity expected
and found in UPC events is to be attributed to the presence of such a colour singlet.
The great importance of UPC processes lies in the capability of probing gluon
distributions at low Bjorken-$x$ at the order of $10^{-5}$.
%

\section{The ALICE detector and the methodology}
\begin{figure}[ht]
\centering
\includegraphics[width=0.8\textwidth]{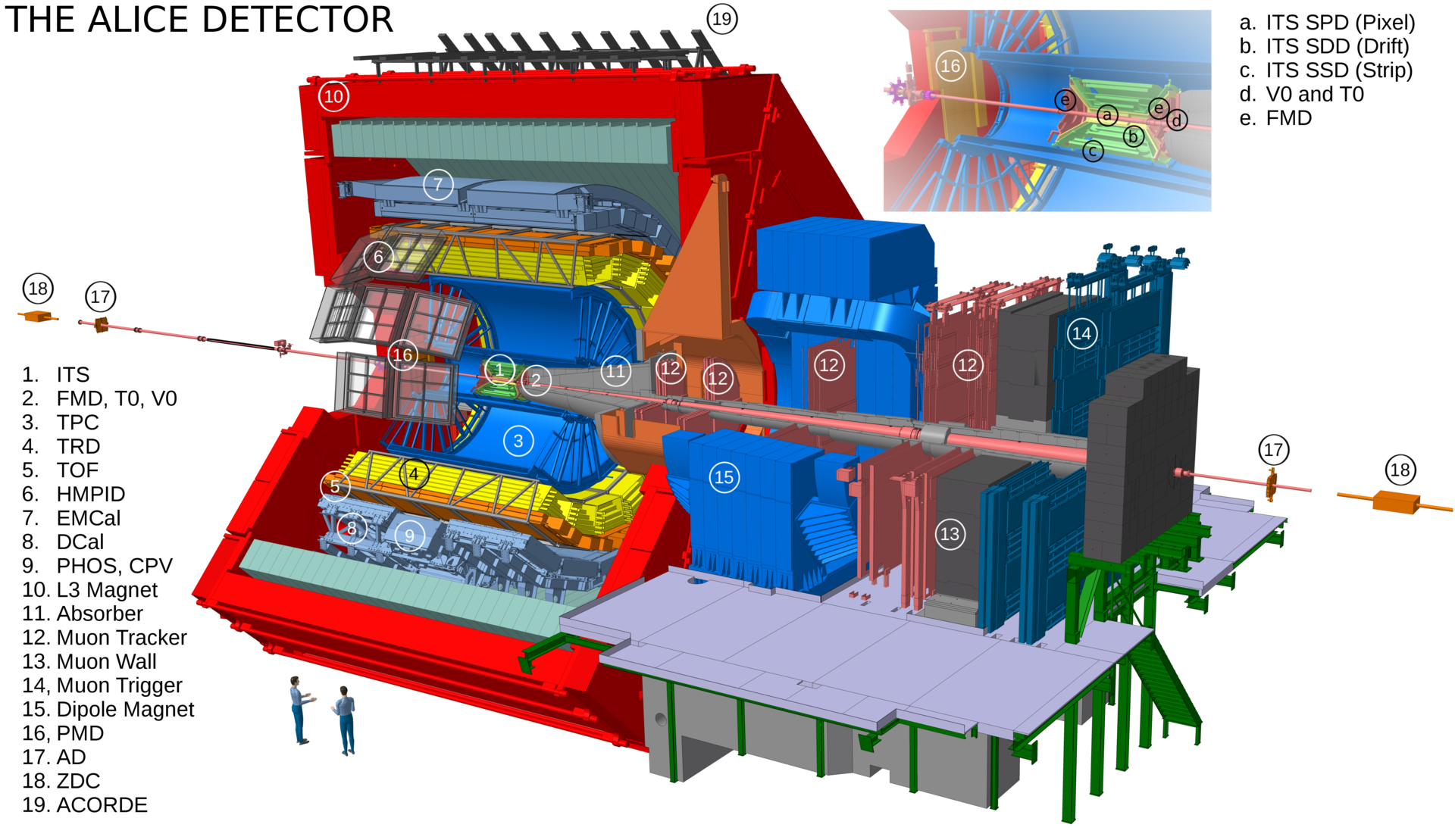}%
\caption{\label{ALICE}The ALICE detector during the LHC Run 2.}
\end{figure}

The ALICE Collaboration has built a detector with excellent tracking and Particle Identification
(PID) capabilities \cite{Collaboration_2008, alice2014performance}, see Fig.~\ref{ALICE}. The detector can be divided in a central
barrel containing e.g. the Inner Tracking System, the Time Projection Chamber
and the Time-Of-Flight System, and a Forward Muon Spectrometer that detects
muons in the forward region $-4.0 < y < -2.5$, where $y$ is the particles'
rapidity. The spectrometer, working in conjunction with the small detectors V0,
AD and ZDC, makes possible the analysis of J/$\psi$ photoproduction in the
forward rapidity region. In the following there will be a distinction between
the A and the C side. The Muon Spectrometer stands on the
C side, while the A side is opposite to it.

UPC events are characterised by a very low multiplicity, in stark contrast to
the extreme multiplicity ALICE usually has to cope with (e.g. 2000 particles
per unit of rapidity in central heavy ion collisions). 
The characteristic feature of UPC events is their topology: only two tracks are
to be expected, in an otherwise empty detector. At forward rapidities, this
means that only  two muons from the J/$\psi$ decay are expected in the Forward Muon Spectrometer, while other detectors are
used as vetoes.

\section{Results}


Results for \textit{coherent} UPC processes are presented here. In this type
of processes, the photon couples directly to the nucleus as a whole.
The coherent component can be separated from the other contributions analysing
the transverse momentum ($p_{\rm T}$) distributions, this is because coherent processes
have a much narrower $p_{\rm T}$-distribution \cite{Abelev:2012ba}.
As such, selecting events with small $p_{\rm T}$ (typically $p_{\rm T} < 0.25$ GeV/$c$)
helps in selecting a sample enriched in coherent J/$\psi$.

\begin{figure}[ht!]
     \begin{center}
        \subfloat[]{
            \label{fig:InvMass}
            \includegraphics[width=0.5\textwidth]{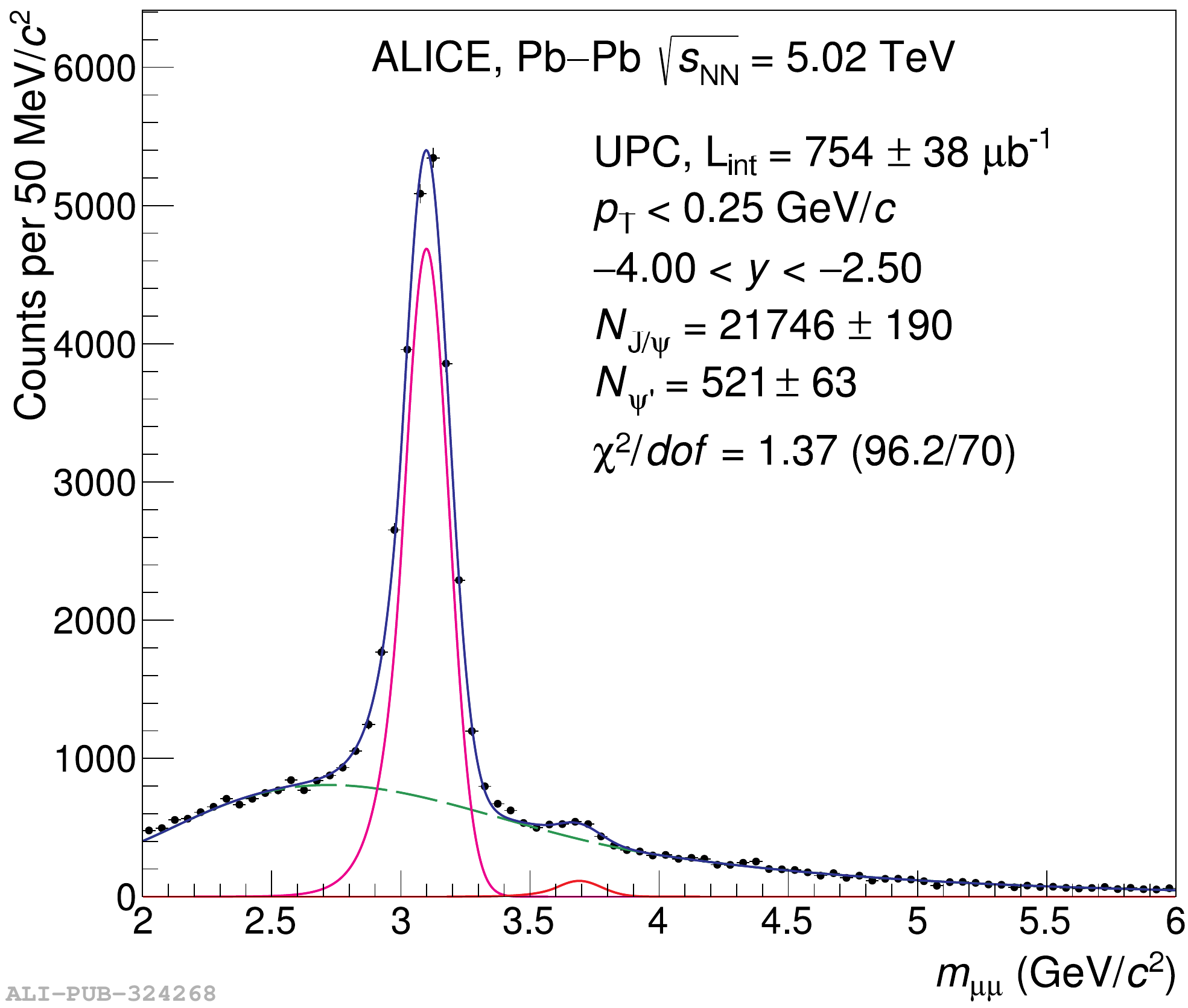}
        }
        \subfloat[]{
            \label{fig:PtDistribution}
            \includegraphics[width=0.5\textwidth]{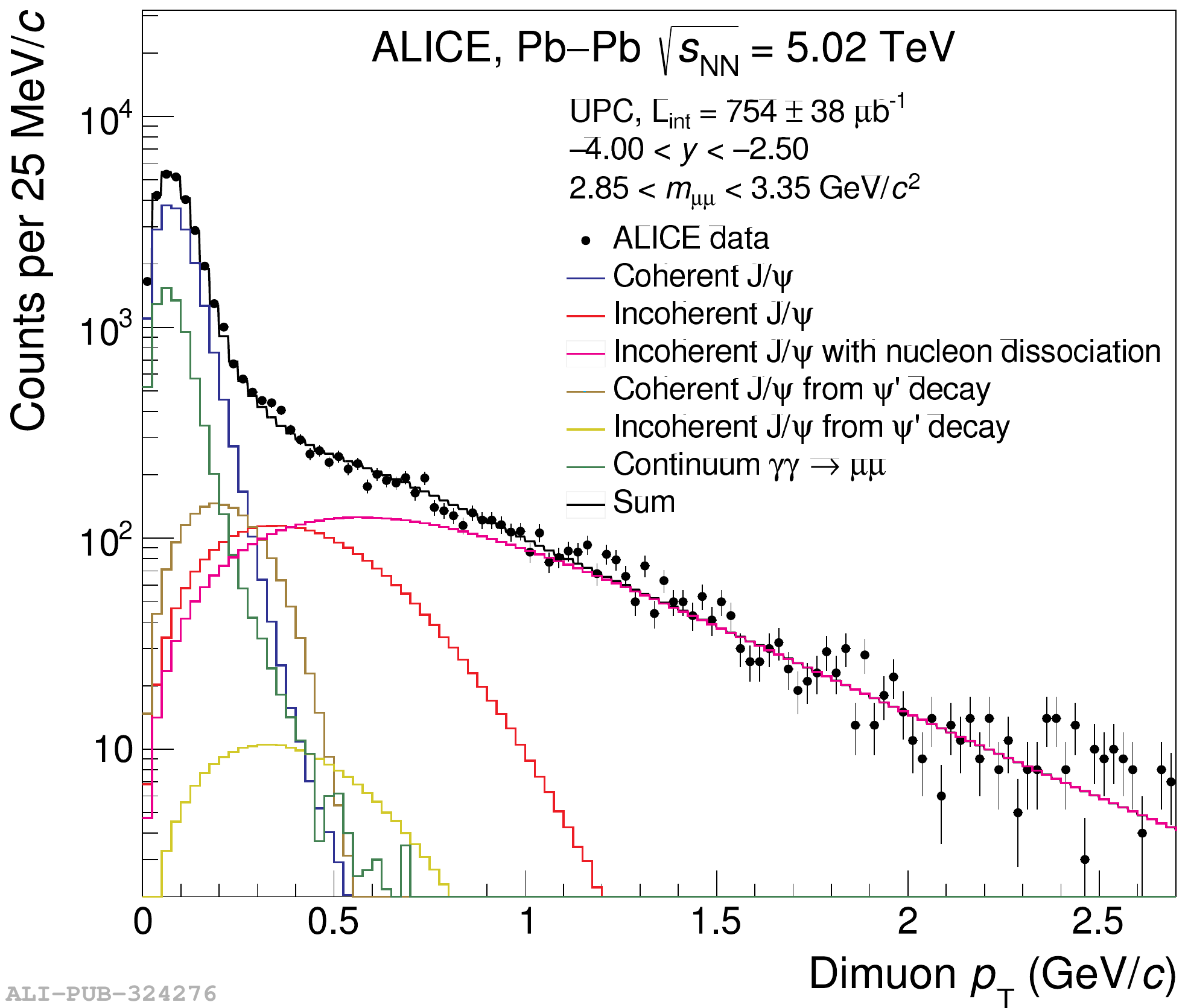}
        }\\ 
    \end{center}
    \caption{a) invariant mass distribution of the dimuons in the forward
             rapidity region, and (b) $p_{\rm T}$-distribution
             for the dimuons within $2.85 < m_{\mu\mu} < 3.35$ GeV/$c^2$.
             Figures taken from \cite{alice2019coherent}.}
   \label{fig:Yield}
\end{figure}

A fit to the invariant mass spectrum for dimuons with $p_{\rm T} < 0.25$ GeV/$c$ is
shown in Fig.~\ref{fig:InvMass}, and was used to extract the ratio of primary coherent
$\psi(2S)$ and J/$\psi$ photoproduction cross sections \cite{alice2019coherent}:
\begin{equation}
  R = \frac{\sigma[\psi(2S)]}{\sigma[\rm{J}/\psi]} = 0.150 \pm 0.018 (stat.) \pm 0.021 (syst) \pm 0.007 (BR)\rm{.}
\end{equation}
Fig.~\ref{fig:PtDistribution} shows the fit to the $p_{\rm T}$-distribution
of the dimuons in the region of the J/$\psi$ mass. This is needed to extract the
contribution of the coherent sample to the whole. The fit is
performed using templates produced with the STARlight event generator \cite{Klein:1999qj, klein2017starlight}.

\begin{figure}[ht]
\centering
\includegraphics[width=0.7\textwidth]{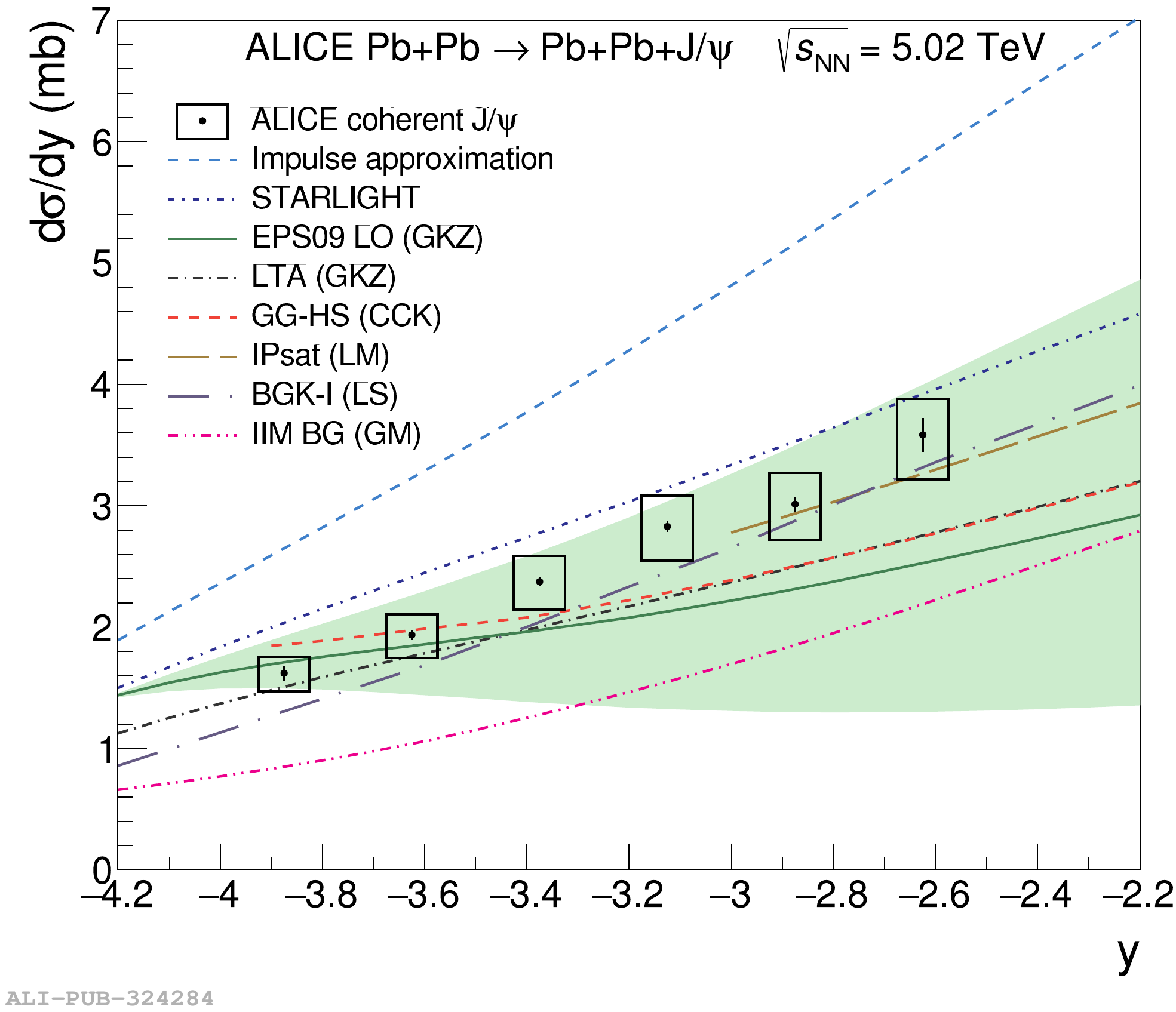}%
\caption{\label{NuclearShadowing} Measured coherent differential cross section of
                                  the J/$\psi$ production in UPC. The error bars
                                  represent the statistical uncertainties and
                                  the boxes the systematic uncertainties.
                                  Figure taken from \cite{alice2019coherent}.}
\end{figure}

Finally, Fig.~\ref{NuclearShadowing} shows the measured differential cross section
of the coherent J/$\psi$ photoproduction in the full range of the considered
rapidity acceptance, together with predictions from different models
\cite{Lappi:2013am, Cepila:2016uku, Cepila:2017nef, Santos:2014zna, Goncalves:2014wna}.
It is shown that the upper variation of the EPS09 (GKZ) model \cite{Guzey:2016piu} with moderate gluon
shadowing gives a fair description of the data, while the impulse approximation model (which does not
consider gluon shadowing at all) is disfavoured. The best agreement is provided by
the BGK-I (LS) \cite{Luszczak:2019vdc} prediction.

\section{Conclusion}
The first measurement of coherent J/$\psi$ photoproduction in UPC Pb--Pb collisions
at $\sqrt{s_{\rm NN}} = 5.02$ TeV with the complete Run 2 dataset has been presented here.
UPC are sensitive to the gluon distribution in nuclei at Bjorken-$x$ at the order
of $10^{-2}$ and $10^{-5}$.
The measured value of $R$ is consistent with the ratio in photon-proton collisions.
The comparison of the measured cross section with the impulse approximation indicates the existence of moderate shadowing.

\ack
The author acknowledges financial support from the School of Physics and Astronomy
at the University of Birmingham and the STFC.

\section*{References}
\bibliographystyle{iopart-num}
\bibliography{JPCSLaTexGuidelines.bbl}

\end{document}